# Vibrational spectra of copper polysilicate, CuSiO$_3$

**Marco Meibohm and Hans Hermann Otto** *

Fachabteilung Materialwissenschaftliche Kristallographie, TU Clausthal
Adolph-Roemer-Straße 2A, D-38678 Clausthal-Zellerfeld, Germany

**Wolfgang Brockner**

Institut für Anorganische und Analytische Chemie, TU Clausthal
Paul-Ernst-Straße 4, D-38678 Clausthal-Zellerfeld, Germany

**Key words:** Copper polysilicate, CuSiO$_3$, Raman and FIR/IR spectra

PACS: 63.20.-e, 63.22, 75.50Ee, 78.30.-j

This is a representation of the first *Raman* and IR / FIR spectra for othorhombic copper polysilicate, CuSiO$_3$, measured at room temperature on polycrystalline samples and a comparison of the optical phonons with those for the spin-*Peierls* compound CuGeO$_3$. CuSiO$_3$ represents a further example of a quasi-one-dimensional spin = ½ antiferromagnetic *Heisenberg* chain system. A mode assignment of the silicate is given. From the analysis of the *Davydov* doublets a reduced intralayer-to-interlayer bond strength of the silicate in comparison to the germanate is obtained allowing for different magnetic response at low temperature.

## I. INTRODUCTION

Low-dimensional quantum spin systems are of considerable theoretical and experimental interest. Some years ago a spin dimerization transition (spin-*Peierls* transition), which is one of the quantum phenomena in a S = ½ antiferromagnetic *Heisenberg* linear chain, was observed in inorganic compounds and extensively studied in the quasi-one-dimensional (1*D*) copper polygermanate, CuGeO$_3$ [1].
The crystal structure of copper polygermanate consists of two types of chains which running down the shortest translation period: *einer* single chains of GeO$_4$ tetrahedra are connected by chains of edge-sharing CuO$_{4+2}$ 'octahedra' [2,3], which are in reality due to the *Jahn-Teller* effect [4] strongly elongated and distorted tetragonal dipyramids. Because of this the magnetic chains actually are formed by ladders of CuO$_2$ squares.

Until now CuGeO$_3$ was the prototypic compound showing such tetrahedral single chains with a repeat sequence of only one GeO$_4$ unit. The substitution of Si$^{4+}$ for Ge$^{4+}$ in copper polygermanate according to CuGe$_{1-x}$Si$_x$O$_3$ was only possible in the limited concentration range x ≤ 0.5 for polycrystals by hydrothermal synthesis [5,21], and the existence of pure CuSiO$_3$ was considered non-existent [6,7]. However, we succeeded in the preparation of polycrystalline CuSiO$_3$, isotypic to CuGeO$_3$, by thermal decomposition of the mineral dioptase, Cu$_6$Si$_6$O$_{18}$·6H$_2$O, and investigated its crystal and magnetic structure [8-11]. Below $T_N$ = 8 K a second order phase transition to a long-range antiferromagnetic *Néel* state order was observed, in contrast to the spin-*Peierls* transition at $T_{Sp}$ = 14 K for CuGeO$_3$.

Strong bonds in the *ab*-plane form zigzag layers in the crystal structure (Fig. 1), but the layer-like character is less pronounced for the silicate than for the germanate, because its interlayer distance is reduced about 3.6% and the intralayer chain distance enlarged about 3.5% in comparison to the germanate. Lattice parameters and CuO$_2$ chain separations of both compounds are compared in Table 1 and individual bond lengths summarized in Table 2 to complete the data used for the further analysis [9-12].

Here we report on a first investigation of the room temperature vibrational spectra of a polycrystalline CuSiO$_3$ sample and compare these measurements with that for CuGeO$_3$ using previously reported data as well as our own.

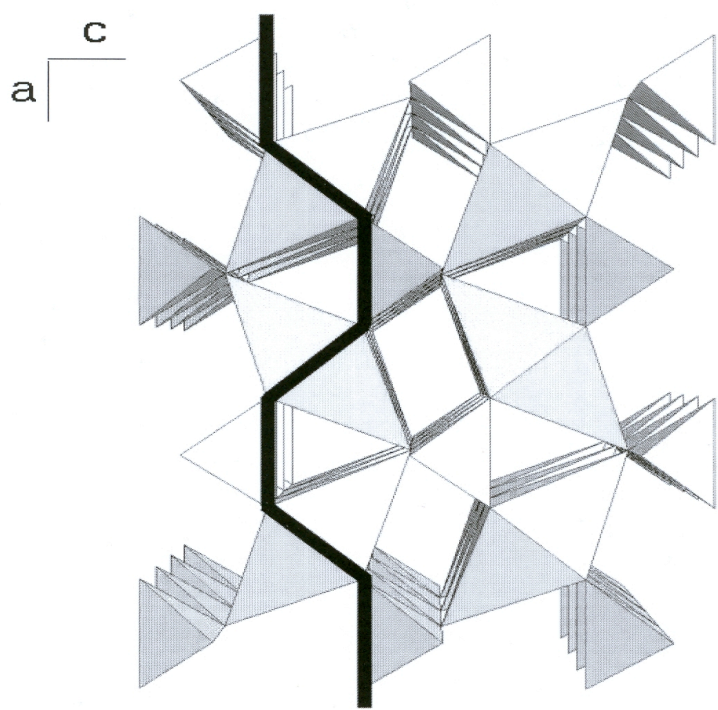

**Figure 1.** Crystal structure projection of CuSiO$_3$ along the chain axis. Zigzag layers in the *ab* plane are depicted by the thick black line.

**Table 1.** Lattice parameters, unit-cell volume and separations of CuO$_2$ chains for isotypic CuSiO$_3$ and CuGeO$_3$. Space group *Pmma* (No 51, standard setting).

| Compound | *a* (Å) | *b* (Å) | *c* (Å) | *V* (Å$^3$) |
|---|---|---|---|---|
| CuSiO$_3$ | 8.7735(11) | 2.8334(4) | 4.6357(6) | 115.24(5) |
| CuGeO$_3$ | 8.4749(3) | 2.9431(1) | 4.8023(2) | 119.78(2) |

**CuO$_2$ chain separations**

| | *a*/2 (Å) (intralayer) | *c* (Å) (interlayer) |
|---|---|---|

**Table 2.** Bond lengths for CuSiO$_3$ and CuGeO$_3$.

| Compound | Atoms | Bond length (Å) | | Reference |
|---|---|---|---|---|
| **CuSiO$_3$** | Cu – O(1) | 2.926(7) | x 2 | [9] |
| | Cu – O(2) | 1.941(4) | x 4 | |
| | Si – O(1) | 1.640(7) | x 2 | |
| | Si – O(2) | 1.582(7) | x 2 | |
| **CuGeO$_3$** | Cu – O(1) | 2.7549(8) | x 2 | [12] |
| | Cu – O(2) | 1.9326(7) | x 4 | |
| | Ge – O(1) | 1.7730(8) | x 2 | |
| | Ge – O(2) | 1.7322(10) | x 2 | |

## II. EXPERIMENTAL

The Raman and IR spectra of CuSiO$_3$ and CuGeO$_3$ were measured on polycrystalline samples. The copper polysilicate specimen has been obtained by thermal decomposition of the mineral dioptase, Cu$_6$Si$_6$O$_{18}$·6H$_2$O, as a multi-phase educt containing about 76 wt-% CuSiO$_3$, 14 wt-% CuO (tenorite) and 10 wt-% amorphous SiO$_2$ [9]. Single-phase powder of CuGeO$_3$ has been synthesized by sintering of pellets of a thoroughly homogenized CuO and GeO$_2$ mixture at 1200 K and a repeated grinding and sintering procedure.

The Raman spectrum of CuSiO$_3$ was recorded with a *DILOR* LabRAM spectrometer, using a Nd-YAG-laser with frequency doubler ($\lambda$ = 532 nm) as an exiting source, and the spectrum of CuGeO$_3$ with a Raman FRA 106 module attached to a *BRUKER* IFS 66v interferometer, using once again a Nd-YAG-laser, but with $\lambda$ = 1064 nm. Infrared spectra of both compounds were obtained with the above mentioned interferometer using CsI pellets of the powdered samples. CsI has a refractive index of about 1.74 very close to the mean refractive index of CuSiO$_3$, which is calculated from specific refractivities of oxides giving n = 1.75 [27].

The IR and Raman spectra of the minor phases in the phase mixture (CuO as tenorite and amorphous SiO$_2$) were also recorded with the same experimental equipment in order to estimate the possible distortion of the CuSiO$_3$ spectra and are given by Meibohm (1999) [10]. Only few strong bands were observed (IR: 470 and 1108 cm$^{-1}$ for amorphous SiO$_2$ and 530 cm$^{-1}$ for CuO; Raman: 450 cm$^{-1}$ for amorphous SiO$_2$ and 300 cm$^{-1}$ for CuO) that would be in a position to introduce a marginal distortion of the main phase spectra. Only the background of the *Raman* spectrum near 300 cm$^{-1}$ as well as the left (high frequency) shoulder of the IR band near 500 cm$^{-1}$ may be influenced by the presence of the minor phases.

## III. FACTOR GROUP ANALYSIS

For the factor group analysis (f.g.a.) the unit-cell, containing two non-translationally equivalent CuSiO$_3$ units and thus n = 10 atoms, were adapted to the standard setting of the *Pmma* space group in order to apply the tables given by Rousseau [13] and Adams [14] and to compare with the data already reported for CuGeO$_3$ [15-22]. From the total number of vibrations ($N_T$ = 3n = 30) the 3 acoustic and 2 silent modes were subtracted, leaving 25 remaining optic modes, 12 of which are *Raman*-active and 13 IR-active modes. The results are summarized in Table 3 displaying the active modes in frames. The alternative space group *P2$_1$2$_1$2$_1$* recently proposed for CuGeO$_3$ in the room temperature phase could not be proved for CuSiO$_3$ [22-24].

**Table 3.** Factor group analysis for orthorhombic CuSiO$_3$ (space group $D_{2h}$–*Pmma*, No. 51).

| Species | Number of modes | | | | | | | | Activity | |
|---|---|---|---|---|---|---|---|---|---|---|
| | Cu | Si | O$_{br}$ | O$_t$ | | | | | | |
| $D_{2h}$ | $C_{2h}$ | $C_{2v}$ | $C_{2v}$ | $C_s$ | $N_T$  $N_A$  $N_O$  $R_y$ | | | $N_C$ | Raman | IR |
| $A_g$ |  | 1 | 1 | 2 | 4 | 1 | 3 | | xx, yy, zz | |
| $B_{1g}$ |  |  | 1 | 1 |  | 1 |  | | xy | |
| $B_{2g}$ |  | 1 | 1 | 2 | 4 | 1  1 |  | 2 | xz | |
| $B_{3g}$ |  | 1 | 1 | 1 | 3 | 1 |  | 2 | yz | |
| $A_u$ | 1 |  |  | 1 | 2 | 1 | 1 | | silent | |
| $B_{1u}$ | 2 | 1 | 1 | 2 | 6 | 1 | 2  3 |  | E∥z | |
| $B_{2u}$ | 1 | 1 | 1 | 1 | 4 | 1 | 1 | 2 | E∥x | |
| $B_{3u}$ | 2 | 1 | 1 | 2 | 6 | 1 | 2  1 | 2 | E∥y | |
| Σ | 6 | 6 | 6 | 12 | 30 | | | | | |

$N_T$ = total number, $N_A$ = acoustic branch, $N_O$ = optic branch, $R_y$ = chain librations, $N_C$ = coupled vibrations of two chains, $O_t$ = terminal oxygen atom, $O_{br}$ = bridging oxygen atom.

The following are features of the analysis that will make the mode assignment easier: a) the copper atoms do not contribute to the *Raman* spectrum; b) coupling between the two non-translationally equivalent silicate chains of $C_{2v}$ site symmetry in the unit-cell with $D_{2h}$ symmetry leads to *Davydov* pairs [25]; the chain modes split into crystal modes of type $A_g$ – $B_{1u}$, $B_{2g}$ – $B_{3u}$ and $B_{3g}$ – $B_{2u}$, respectively [16]. The obtained doublets involving both *Raman*- and IR-active modes are indicated in Table 3 by vertical lines.

## IV. RESULTS and DISCUSSION

The room temperature Raman and infrared spectra of both $CuSiO_3$ and $CuGeO_3$ are shown in Figs. 2 and 3, and the extracted vibrational frequencies are summarized in the Table 4 along with their intensities and the mode assignments. Even though excellent spectra on $CuGeO_3$ single crystals are available in the literature, our measurements on powder samples are still given to compare the powder spectra. At a rough estimate the spectra of the silicate shift on average by a factor of $F = F_{m*} \cdot F_{bl} = 1.23$ to higher frequencies in comparison to that of the germanate. This factor accounts for the frequency ratio of the internal vibrations of the tetrahedral groups, but may be applied to the whole $CuSiO_3$ spectra. In this way the mode assignment for the germanate, confirmed by a normal coordinate analysis [16], can be applied to the silicate spectra.

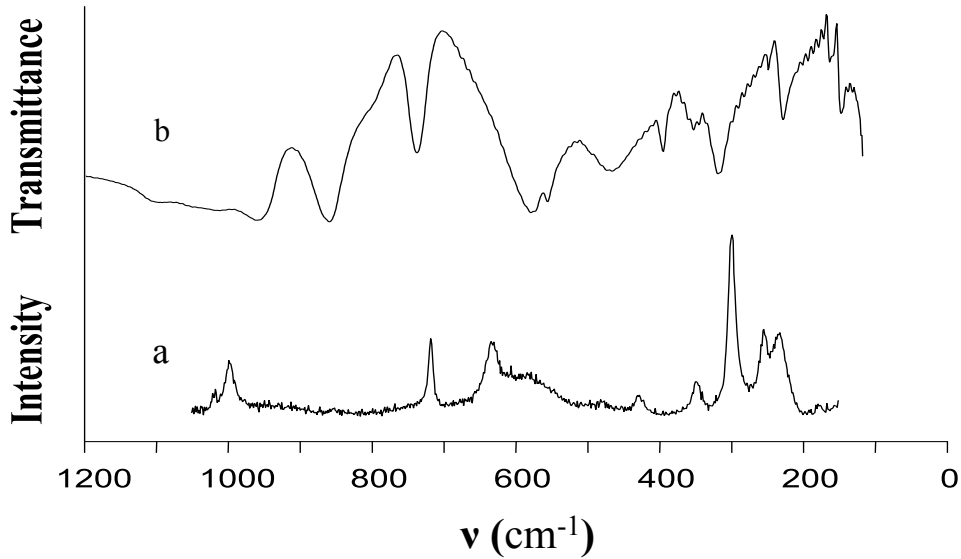

**Figure 2.** *Raman* (a) and IR / FIR (b) spectra of $CuSiO_3$ at room temperature.

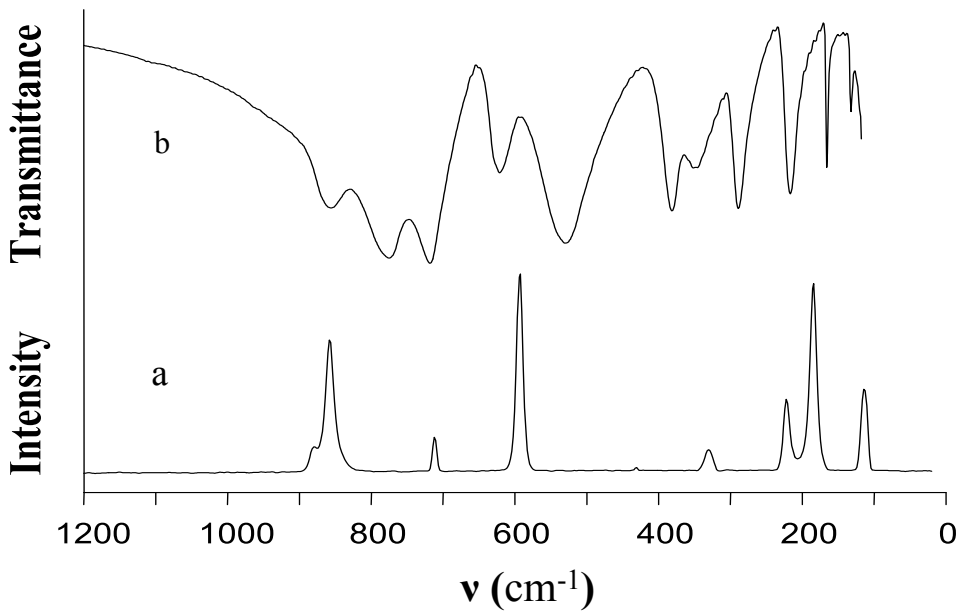

**Figure 3.** *Raman* (a) and IR/FIR (b) spectra of $CuGeO_3$ at room temperature.

**Table 4.** Vibrational frequencies (cm$^{-1}$) and mode assignment for CuSiO$_3$ and CuGeO$_3$.

| Cu[GeO$_3$] Raman | Cu[GeO$_3$] IR | Species | Cu[SiO$_3$] Raman | Cu[SiO$_3$] IR |
|---|---|---|---|---|
| 880 w | | B$_{2g}$ | 1019 w | |
| 858 s | | A$_g$ | 998 m | |
| | 857 s | B$_{1u}$ | | 1092 m |
| | 776 vs | B$_{3u}$ | | 959 s |
| | 718 vs | B$_{2u}$ | | 859 vs |
| 712 w | | B$_{3g}$ | 854 vw | |
| | 619 m | B$_{1u}$ | | 737 s |
| 592 vs | | A$_g$ | 719 m | |
| | | | 634 m ? | |
| | | | 590 m,b ? | |
| | 528 vs | B$_{2u}$ | | 576 vs |
| | 478 m,sh | B$_{1u}$ | | 558 m |
| 430 vw | | B$_{2g}$ | 480 vw | |
| [411] vw | | B$_{3g}$ | — | |
| [388] vw | | B$_{1g}$ | — | |
| | 380 s | B$_{3u}$ | | 465 m,b |
| | 347 m | B$_{1u}$ | | |
| 330 w | | A$_g$ | 429 w | |
| | 289 s | B$_{3u}$ | | 395 w |
| | | | | 353 vw ? |
| 222 m | | B$_{2g}$ | 349 w | |
| | 215 s | B$_{3u}$ | | 317 m |
| 184 vs | | A$_g$ | 300 vs | |
| | 166 m | B$_{2u}$ | | 227 m |
| | 132 w | B$_{1u}$ | | 160 w |
| 113 m | | B$_{2g}$ | 255 m | |
| 110 | | B$_{3g}$ | 233 m | |
| | [48] vw | B$_{3u}$ | | *) |

vs = very strong, s = strong, m = medium, w = weak, vw = very weak, sh = shoulder, b = broad, O$_t$ = terminal oxygen atom, O$_{br}$ = bridging oxygen atom, vw? = very weak peak on the shoulder of a medium peak, — *Davydov* pairs, *) outside of measurement range. Frequencies in squares were not observed in the powder spectra, but have been reported for single crystal spectra. Broad *Raman* bands of CuSiO$_3$ with question marks may be caused by surface effects.

In particular is $F_{m*} = (m*_{germanate} / m*_{silicate})^{1/2} = 1.13$, where m* is the reduced mass, and $F_{bl} = <d>_{germanate} / <d>_{silicate} = 1.08$; the mean bond lengths (bl) were applied as a measure of the stiffness of the compounds with $<d> = 1.747$ Å for the germanate and $<d> = 1.611$ Å for the silicate (Table 2).

In Fig. 4 the frequency quotient of silicate to germanate modes versus the germanate frequency is displayed and affords some information about the source of the vibrations. In this drawing three regions of different slope can be made out, beginning with a low frequency region below 250 cm$^{-1}$ (350 cm$^{-1}$ for the silicate) originating from Cu and Ge/Si atom vibrations, then a region between 250 cm$^{-1}$ and 550 cm$^{-1}$ (350 to 600 cm$^{-1}$) with Cu–O and Ge/Si–O vibrations, and finally the region above 550 cm$^{-1}$ showing modes with a dominant contribution of oxygen vibrations. For the minimum values of the curve it is allowed to neglect $F_{m*}$ thus obtaining a value, which actually reflects the force constant ratio. A distinction is also made between *Raman* and IR data in order to demonstrate the difference of their low frequency modes. Because the copper atom does not contribute to the *Raman* spectrum the large Si to Ge mass ratio is decisive for the large frequency ratio, while for the IR spectrum the contribution of Cu to the reduced mass level out the frequency difference between germanate and silicate.

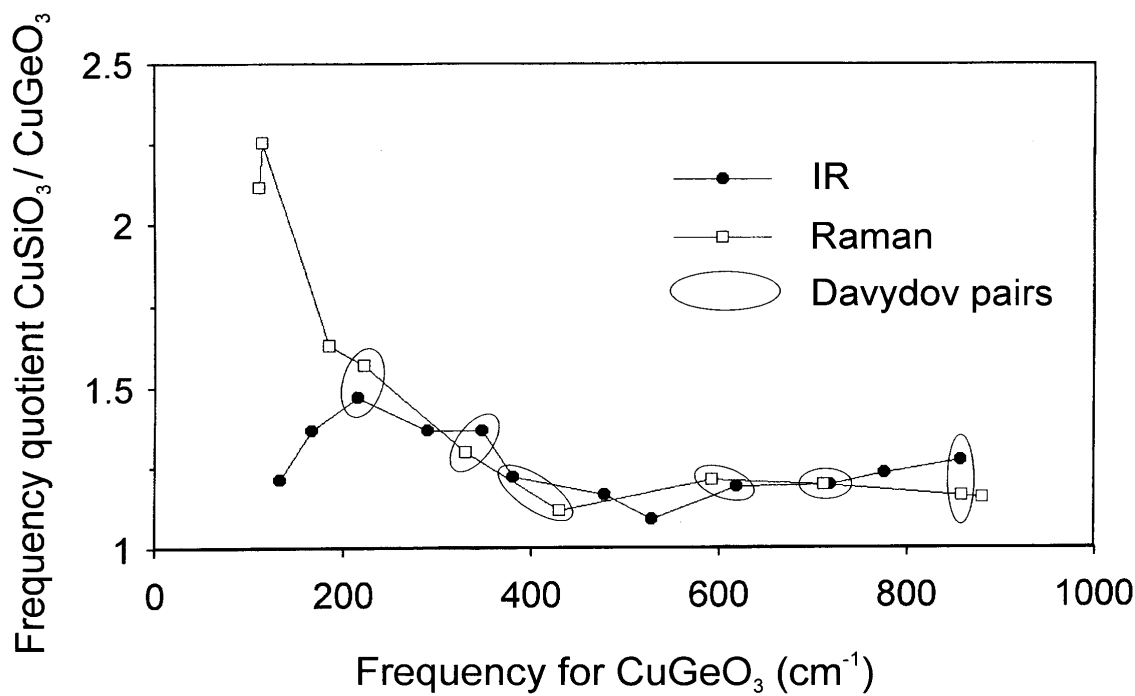

**Figure 4**. Frequency ratio of silicate to germanate modes versus the germanate frequency. Raman data are depicted as empty squares, IR data as filled circles. Davydov pairs are indicated by elliptical frames.

The number of modes found in the spectra is in agreement with the space group symmetry and f.g.a. taking into account the following points. Three very weak bands in the vibrational spectra of CuGeO$_3$, given in Table 4 in squares, which have been observed in polarized single crystal spectra [16], are absent in the spectra of powder samples of both CuGeO$_3$ and CuSiO$_3$ due to very low intensity. For the silicate, the peak with the lowest Raman frequency obviously lies outside of the measurement range. A very broad band at 465 cm$^{-1}$ in the IR-spectrum of CuSiO$_3$ is interpreted as a double peak. Two very broad bands in the *Raman* spectrum of CuSiO$_3$, showing no counterpart in the CuGeO$_3$ spectrum, may be caused by surface effects due to the very small crystallite size, which is concluded from the broad X-ray diffraction peaks observed [9]. Finally, a very weak side band at 353 cm$^{-1}$ observed in the IR spectrum of CuSiO$_3$ may belong to interference fringes caused by the parallel alignment of reflecting surfaces within the pressed CsI disk.

Once the correct mode assignment has been found, it is sufficient to apply to the silicate the results of the normal coordinate analysis for the germanate, already given by Popović et al. [16], in order to find out what the origin of an individual vibration is.

Weiden et al. [21] observed defect modes in the *Raman* spectra of Si-doped $CuGeO_3$. All defect modes belong to the $A_g$ type. For instance, a strong mode at 592 cm$^{-1}$ of pure $CuGeO_3$ corresponds evidently to a defect mode at 670 cm$^{-1}$. This mode is assumed to develop into the 719 cm$^{-1}$ mode of pure $CuSiO_3$ (Table 4), because the frequency shift is stronger for high Si content than the linear extrapolation obtained from lower Si doping would suggest. This mode arises from $O_1$ oxygen atom bond-stretching vibration along the c axis [16]. Another defect mode at 205 cm$^{-1}$, originating from the very strong 184 cm$^{-1}$ mode shifts to the very strong mode at 300 cm$^{-1}$ of the pure silicate.

With regard to the crystal symmetry, the spectroscopic data as well as X-ray diffraction results are supported by a further experiment. According to NQR measurements of Cu in $CuSiO_3$ there exists, even down to 4.2 K, only one Cu site [26].

Following Popović et al. [16], we can analyse the chain mode doublets (*Davydov* pairs) for $CuSiO_3$, like that for $CuGeO_3$, as the result of the vibration of a pair of weakly coupled identical oscillators by the relation $\nu_\pm = (\nu_0^2 \pm \Delta^2)^{1/2}$, where $\nu_0$ is the frequency of the isolated oscillators and $\Delta^2$ is proportional to the coupling force constant. Then $(\nu_0/\Delta)^2$ gives a measure of the ratio of intralayer-to-interlayer bond strength. Table 5 summarizes the phonon doublet analysis ($\nu_+ = \nu_{Raman}$, $\nu_- = \nu_{IR}$), which yields the mean value of $(\nu_0/\Delta)^2$ for the silicate about 20% lower than for the germanate [16] in accordance with the changed chain separations that have been derived from the crystal structure analysis (Table 1).

**Table 5.** Analysis of *Raman* and infrared active phonon doublets for $CuSiO_3$. Frequencies $\nu$ and $\Delta$ are given in cm$^{-1}$.

| $\nu_{Raman}$ | $\nu_{IR}$ | $\nu_0$ | $\Delta$ | $(\nu_0/\Delta)^2$ |
|---|---|---|---|---|
| $A_g$ | $B_{1u}$ | | | |
| 429 | 465 *) | 447 | 128 | 12 |
| 719 | 737 | 728 | 115 | 40 |
| 998 | 1092 | 1045 | 313 | 11 |
| $B_{2g}$ | $B_{3u}$ | | | |
| 349 | 317 | 333 | 105 | 10 |
| 480 | 465 *) | 473 | 85 | 31 |
| $B_{3g}$ | $B_{2u}$ | | | |
| 853 | 859 | 856 | 72 | 143 |
| | | | $\langle(\nu_0/\Delta)^2\rangle =$ | 41 |

*) This broad band is interpreted as a double peak.

The more 3*D*-like character of the silicate may be responsible for the different spin-phonon coupling in comparison to the germanate leading at low temperature to a long-range antiferromagnetic *Néel* state order instead of a spin-*Peierls* order.

## V. SUMMARY and OUTLOOK

The vibrational spectra of orthorhombic $CuSiO_3$ from powder samples are presented. The observed modes are assigned based on single crystal results for $CuGeO_3$ previously given. A chain mode analysis of *Davydov* pairs suggested that the intralayer-to-interlayer bond strengths for the silicate are about 20% lower than for the germanate, reflecting the different magnetic behavior of the silicate and the germanate.

Single crystals of $CuSiO_3$ are required to improve the obtained vibrational spectra and extend the measurement to low temperatures giving the experimental basis for a meaningful lattice dynamical calculation. The negligible mismatch between certain lattice parameters of tenorite (CuO) and $CuSiO_3$ suggests the possibility of a metastable formation of $CuSiO_3$ by way of a topotactically induced reaction on the tenorite surface [9]. It is proposed to use a tenorite substrate or at least a tenorite buffer layer for an epitaxial growth of well crystallised copper polysilicate.